\documentclass{article}

\usepackage{PRIMEarxiv}
\usepackage{siunitx}
\usepackage[utf8]{inputenc} % allow utf-8 input
\usepackage[T1]{fontenc}    % use 8-bit T1 fonts
\usepackage{hyperref}       % hyperlinks
\usepackage{url}            % simple URL typesetting
\usepackage{booktabs}       % professional-quality tables
\usepackage{amsfonts}       % blackboard math symbols
\usepackage{nicefrac}       % compact symbols for 1/2, etc.
\usepackage{microtype}      % microtypography
\usepackage{lipsum}
\usepackage{fancyhdr}       % header
\usepackage{graphicx}       % graphics
\graphicspath{{media/}}     % organize your images and other figures under media/ folder

\title{A Synopsis of Stent Graft Technology Development
\thanks{\textit{\underline{Citation}}: 
\textbf{Authors. Title. Pages.... DOI:000000/11111.}} 
}

\author{
  Umme Hafsa Momy \\
  Florida International University \\
  Miami\\
  \texttt{umomy001@fiu.edu}
}

\begin{document}
\maketitle

\begin{abstract}
Coronary artery disease (CAD) is the predominant cause of mortality and morbidity across the globe. Over the past few decades, treatments for CAD have witnessed dramatic evolution, with percutaneous coronary intervention (PCI) with stenting taking precedence over bypass surgery as the primary revascularization strategy. This paper delivers an extensive overview of the significant progress in coronary stent technology, tracing back to the inaugural coronary angioplasty in 1977. Early trailblazers like Werner Forssmann, Charles Dotter, and Andreas Gruentzig laid the groundwork for interventional cardiology. The introduction of bare metal stents (BMS) in the late 1980s offered solutions to the limitations of balloon angioplasty, such as acute vessel closure and restenosis. However, BMS had its own set of challenges. Consequently, the early 2000s saw the emergence of first-generation drug-eluting stents (DES), utilizing sirolimus and paclitaxel, offering significant reductions in restenosis compared to BMS. Despite their success, safety concerns such as very late stent thrombosis arose. Innovations continued with the second-generation DES, featuring advanced stent platforms and biocompatible polymers, ensuring enhanced long-term results. The most recent advancement has been the bioresorbable vascular scaffolds (BVS), which are designed to resorb over time, eliminating the need for a long-term metallic implant. Throughout this journey, clinical trials played a pivotal role in validating the efficacy of each stent generation. While there have been remarkable improvements in reducing restenosis and other adverse events, challenges like optimizing regulatory approval pathways, stent selection, and minimizing risks associated with thrombosis and restenosis persist. The future holds promise for more individualized stenting strategies, tailored to specific patient and lesion profiles. This review not only traces the rapid evolution of coronary stent technology but also underscores its transformative impact on patient care and outcomes.

\end{abstract}

% keywords can be removed
\section{keywords}
\textit{Coronary artery disease, Percutaneous coronary intervention, Coronary stenting, Bare metal stents,Angioplasty,
Revascularization Drug-eluting stents, Bioresorbable scaffolds, Restenosis, Stent thrombosis.
}

%%%%%%%%%%%%%%%%%%%%%%%%%%%%%%%%%%%%%%%%%%%%%%%%%%%%%%%%%%%
%%%%%%%%%%-- Introduction 
%%%%%%%%%%%%%%%%%%%%%%%%%%%%%%%%%%%%%%%%%%%%%%%%%%%%%%%%%%%%%%

\section{Introduction}
Coronary artery disease (CAD) is a leading cause of mortality worldwide, responsible for over 9 million deaths in 2016 \cite{naghavi2017global}. Percutaneous coronary intervention (PCI) with stenting has become the most commonly performed revascularization procedure for obstructive CAD \cite{dehmer2012contemporary}. Since the first coronary balloon angioplasty was performed in 1977 \cite{gruntzig1979nonoperative}, there has been rapid development and evolution of stent technology over the past decades. Bare metal stents (BMS) were introduced in the late 1980s to overcome limitations of balloon angioplasty like acute vessel closure and restenosis \cite{serruys1994comparison,fischman1994randomized}. First-generation drug-eluting stents (DES) emerged in the early 2000s to further reduce restenosis rates compared to BMS \cite{moses2003sirolimus,stone2004polymer}. However, concerns emerged regarding very late stent thrombosis with first-generation DES \cite{camenzind2007response}. This spurred development of second-generation DES with novel stent platforms and polymers to improve long-term safety and efficacy \cite{stefanini2011long}. Most recently, bioresorbable vascular scaffolds (BVS) have been introduced as a transient scaffold to provide short-term vessel support and drug delivery without leaving a permanent metallic implant \cite{ormiston2008bioabsorbable}.

In this paper, we provide a comprehensive overview of the major developments in coronary stent technology over the past 30 years since the introduction of BMS. We summarize the evolution of stent designs and materials from early BMS to modern second-generation DES and BVS. We review clinical data from landmark trials comparing different stent generations and technologies. Finally, we discuss future directions for coronary stenting with a focus on optimizing patient outcomes and minimizing adverse events like restenosis and stent thrombosis. This synopsis of the rapid growth of stent technology illuminates how each new stent generation aimed to incrementally improve upon limitations of its predecessors.

\section{Method}

\subsection{Werner Forssmann (1904–1979)}

Werner Forssmann was the first to venture into the coronary arteries, because the human heart has
always fascinated him. He invented the notion of giving drugs to the heart with a catheter in 1929, which
was a completely unknown concept. Even though no institution would allow him to attempt such a thing,
he was undeterred and made the decision to proceed anyway without giving much thought to the likely
outcomes. He asked the nurse to numb his own left elbow while she quickly cut, a urethral catheter was
then inserted into his own arm after the vein was unlocked. After more spins, he had Dozen escort him to
the radiology lab in the basement, where he used the catheter inserted into his heart to take
photographs \cite{monagan2007journey,forssmann1997werner,king1998development}.

Although Forssmann was unable to continue his research on cardiac catheterization, At Columbia
University in New York, Andre F. Cournand and Dickinson W. Richards took his idea and developed it
further decades later, in the late 1930s. They improved the method and used it to take meaningful
measurements inside the heart. The catheter technique quickly replaced other techniques as the accepted
method of measuring intracardiac pressure. However, it would be some time before the crown tree and
its tangled branches could be located

\subsection{F. Mason Sones (1919–1985)}

In 1950, F. Mason Sones signed on with the Cleveland Clinic. He was intelligent, tenacious, and stayed in the hospital for the majority of his time.

Sones instructed his colleague to inject a dye shot into the aorta to illuminate it. In his laboratory, Sones placed a diagnostic catheter into the ascending aorta of a young patient on October 30, 1958.The oxygen-free angiographic dye was intended to prevent oxygen delivery and cause ventricular fibrillation, but a small catheter suddenly began to whip around like an uncontrollable garden hose was used to inject all of the dye deeply into the patient's right coronary artery. Nothing like this had ever been tried before. The patient was unharmed throughout, and the procedure produced a precise image of the coronary arteries. Sones proclaimed with triumph, He claimed that they had "just revolutionized cardiology”, after all. Using diagnostic catheters to create incredibly detailed images, he was able to successfully place aortic trunk arteries \cite{prize2020physiology}.

Sones began developing unique tapered-tip catheters with an open end and a mesh just beginning to prevent the catheter from clogging the vessel's shaft and was inspired by the unexpected incident and outcome \cite{monagan2007journey}. It did not take long for coronary angiography to become established as a safe and common test for CAD. A special J-shaped catheter was developed by Melvin P. Judkins, who modified the method to make catheterization of the coronary arteries easier and require less effort.However, it requires extensive practice to get a catheter into the tiny openings of the coronary arteries \cite{cowley2005tribute}.

\subsection{Charles T. Dotter (1920–1985)}
The Director of radiology at the University of Oregon in Portland is the brilliant Charles T. Dotter., developed numerous methods for detecting and treating vascular disease \cite{kinney1996radiologic,doby1984tribute}.

Sven-Ivar Selinger, known for his method of inserting a catheter inside blood vessels placing a catheter inside blood vessels— "Catheter over the wire, needle in, wire in, needle out" had spent some time with him.10 He conducted research using various materials, including Piano wires, guitar strings, and other cables and he produced personal catheters because he believed that catheter technology could be used for more than just diagnosing problems.

Dotter identified a chance and provided her with a novel procedure never done before. 83 years of age diabetic lady with a non-healing toe with gangrene and a foot sore underwent an arteriogram of her left leg on January 16, 1964. Her surgeons had insisted on amputating the foot, believing she was beyond help with her poor circulation. but the lady vehemently declined. He first passed a guidewire through the plaque obstruction before inserting a small-caliber catheter and then wedged the plaque by inserting larger and larger catheters through it. In an instant, the lady's frozen leg warmed up and became hyperemic, seemingly by magic. The woman's ulcer was healed after a few weeks, and the pain was subsided, according to X-rays that showed improved circulation \cite{payne2001charles,radiopaedia_charles_dotter}.

\subsection{Andrew}
In 1969, Andreas Gruentzig joined University Hospital of Zurich after completing his medical undergraduate studies in Heidelberg. Gruentzig was fascinated by the "Doddering" technique and had learned it from a lecture, but he also believed it needed improvement because there was a high potential for vascular harm, plaque falling off and acute distal occlusion due to embolization.

Gruentzig developed the concept of opening a blood vessel with a balloon attached to the catheter's tip. in his wild and revolutionary imagination Except for Michaela and Maria Schlumpf, no one else expressed much interest in or support for his idea in 1972 because it was so outlandish. Since they lacked a laboratory and his research funding, using his kitchen as a a workspace for the following two years, working on his catheters almost every evening with his wife and Schlumpf.

It was difficult to get the catheter's tip fitted with a balloon so that he could blow it up to open a locked container. He could blow it up to unlock a locked container was no easy task, and they encountered many technology issues, including air leaks and numerous balloons that expand asymmetrically or lose their structural integrity. However, he persisted and adjusted. As a result, he experimented with different materials, shapes, and designs, but after hundreds of failures and successes, he began to see a few encouraging outcomes.4

He experimented with his design on diseased arteries taken from cadavers and animal models, and as his methods for making balloons got better, he felt ready to use it on a patient.

It was February 1974, Gruentzig pushed his catheter and inflated the balloon in a man in his 67s whose lower extremity pain rendered him unable. As a result, patient’s pain resolved immediately. After that, he moved his next challenge. While awaiting a chance to use his catheters in a human body, he met with numerous production companies, engineers and the layout of his device was constantly being improved.

\section{ A Simple Angioplasty is Insufficient}

On September 16, 1977, Bachmann didn't show any signs of pain The artery was visible as being open on radiographic images. A significant accomplishment in medicine, it was a big success. In addition to other cases, Gruentzig was successful in reproducing the outcomes, and he made the news. On February 7, 1978, the front page of a Swiss newspaper carried the headline "Medical Sensation: Balloon Treatment Against Heart Attacks \cite{monagan2007journey}.

\begin{figure}[h]
    \centering
    \includegraphics[width=0.7\linewidth]{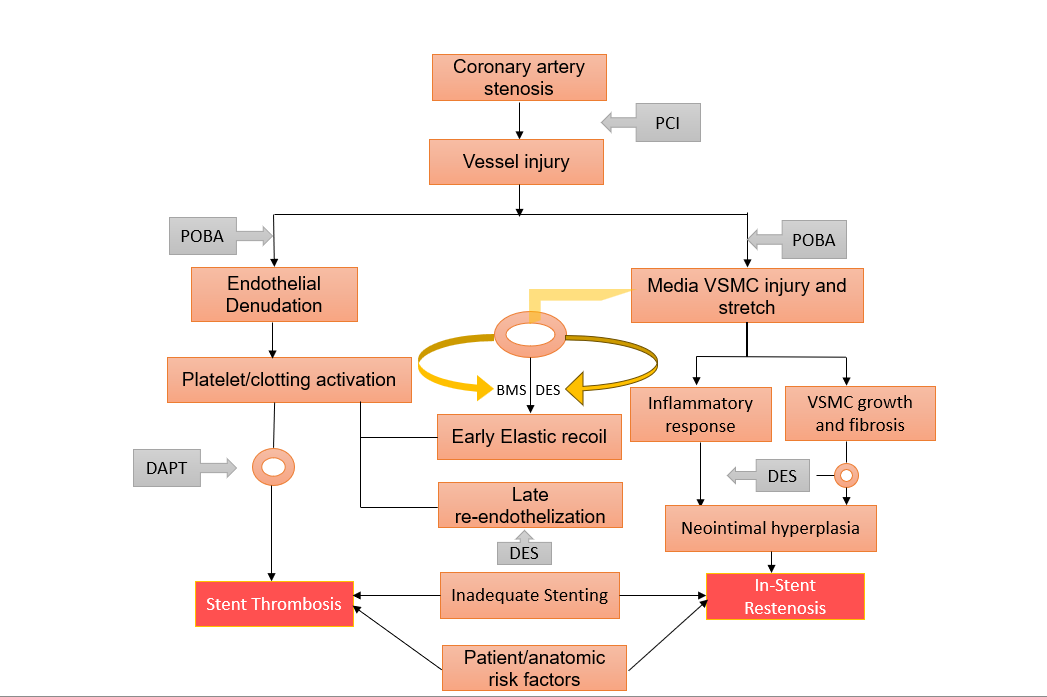} % Adjust the width as necessary
    \caption{Pathophysiology of In-Stent Restenosis and Thrombosis.}
    \label{fig1} % For referencing the figure later in the document
\end{figure}
Geoffrey O. Hartzler was one of the many followers Gruentzig had trained over the years \cite{kahn2004geoffrey}.He shocked the medical community in 1980 by using angioplasty to try to destroy a myocardial infarction (MI) as it developed, which kind of signaled a fresh approach for the procedure. In addition, Hartzler started modifying the stiff end of the catheter to have better curves so that it would slide into place more easily as he grew impatient with Gruentzig’ s invention's flaws.

Hartzler has pushed the boundaries further, making the bold claim that failure of angioplasty is the only cause for bypass surgery Despite the fact that claim merely partially accurate, initial studies contrasting using surgical techniques for angioplasty, for instance the Coronary Angioplasty vs Bypass Revascularization Investigation (CABRI) \cite{participants1995first}, the Bypass Angioplasty Revascularization Investigation (BARI) \cite{bypass1996comparison},Emory Angioplasty vs Surgery Trial (EAST) \cite{bypass1996comparison}, the German Angioplasty Bypass Surgery Investigation (GABI) \cite{bypass1996comparison}, the Randomized intervention Treatment of Angina (RITA) \cite{bypass1996comparison}, showed this in selected patients.

By extruding plaque, simple balloon angioplasty can temporarily increase lumen diameter, but elastic recoil quickly eliminates this gain. Plaque dissection can produce plastic, more permanent changes, but there is a possibility of acute vascular occlusion with this method \cite{serruys1994comparison}.

The inventors of coronary angioplasty were forced to perform these procedures in an active surgical standby mode due to an abrupt occlusion. When balloon-induced intimal denudation and medial tearing occur, the subendothelial matrix is exposed to the blood, which promotes platelet aggregation and thrombosis in the acute phase and chronic negative changes in vascular remodeling (late recoil) and neointimal hyperplasia in the chronic phase. In the first 6 to 9 months, 30 percent to 40 percent of patients experienced an almost total loss of therapeutic benefit due to insufficient initial reinforcement and restenosis (Fig. \ref{fig1}) \cite{fischman1994randomized}.

\section{Key Preclinical Studies}

Preclinical studies in animal models like porcine coronary arteries provided key insights into optimal stent design features. For example, Saxon et al. compared different stent materials and configurations in a pig model. They found tantalum wire stents achieved a larger acute lumen than stainless steel, with less thrombosis risk \cite{saxon2003radiographic}. Schwarzacher et al. tested antithrombotic stent coatings like carbon and heparin in a sheep model, finding reduced acute thrombosis compared to uncoated stents \cite{moustapha2002combined}. These studies guided early stent material and coating selections prior to clinical use. Later preclinical work focused on evaluating safety of stent polymer coatings and kinetic drug release profiles for developing drug-eluting stents \cite{de2004time}.

\section{Stent Development}
In 1986, Ulrich Siegwart of Switzerland and Jacques Puel of France implanted the first stents\cite{sigwart1987intravascular}. Coil stents and slotted tube designs were first implanted after these self-expanding mesh designs in 1987 at Emory University Hospital and So Paulo, Brazil, respectively. With the spread worldwide Angioplasty procedures revealed that the arteries of numerous patients gradually narrowed hours, days, or even months following the operation. As techniques for angioplasty spread across the globe, it became apparent that many patients' arteries gradually narrowed weeks or months after the procedure\cite{volker2002covered,zong2022advances,albohy2023https}. Early research revealed rates of restenosis ranging between twelve and forty-eight percent \cite{serruys1988incidence}.Stents were created that would keep vessels conserve after angioplasty to address these problems. Studies have exhibited improved fast outcomes and longer period without events stenting associated to standard inflatable angioplasty. Restenosis levels were declined by about 10 percent overall \cite{fischman1994randomized,serruys1994comparison}.

\section{Classification of Stent}
The most significant development within transcutaneous coronary revascularization is the development of stents. Restenosis within a stent. is no longer a significant issue with coronary intervention to make stents that can be infused with drugs. Metal stents covered with a polytetrafluoroethylene (PTFE) membrane, known as covered stent-grafts, were developed to prevent restenosis caused by the growth of tissue through the mesh of a stent. An intriguing idea for preventing intraluminal proliferation, sealing degenerated vein grafts and covering coronary artery perforations is the use of stent grafts, which integrate a membrane into a coronary stent.

After balloon dilatation, arterial recoil and restenosis should be avoided by using coronary stents. Bare metal stents (BMS), drug eluting stents (DES), and bioresorbable vascular scaffolds (BRS) are the three major types of stents.

\subsection{Bare metal stents (BMS)}
Bare metal stents (BMS) were the first stents used. These stents can be made thinner, have great mechanical strength and poor flexibility. However, bare metal stents can cause restenosis and can lead to peripheral embolism after being implanted in old vein grafts. A lower restenosis rate, as confirmed by two historical trials issued in 1993 \cite{serruys1994comparison,fischman1994randomized}.Observational studies indicate that the rate of cardiovascular events is increased when stent grafts are used voluntarily in native vessels. However, the strong mechanical support also contributes to neo-intimal hyperplasia. Intravascular ultrasound studies showed that stents required high pressures to fully expand, leading to the development of Dual antiplatelet therapy (DAPT) combining ticlopidine, clopidogrel, and aspirin. In-stent restenosis (ISR), observed in mid- and long-term follow-up, 15 percent to 30 percent of treated lesions was still significantly risky with these stents \cite{hoffmann1996patterns}.

\subsection{Drug-Eluting Stents (DES)}
Drug-Eluting Stents (DES) the next creation of stents. The drug eluted was an antimitotic agent that prevented the growth of SMCs. In the history of interventional cardiology, there has been a third revolutionary paradigm shift. was signaled during 1999 when the first DES was implanted in Brazil by Sousa. However, the possibility of late stent thrombosis (ST) is enhanced by impaired endothelial regeneration and vasomotion.

\subsubsection{First-generation Drug-eluting Stents}
First-generation DESs originally used two antiproliferative medications were sirolimus and paclitaxel. When Cementin published a meta-analysis in 2006, It was discovered that myocardial infarction (MI) and mortality risk were both increased by stent thrombosis (ST) that developed extremely late or tardy \cite{camenzind2007cause}. Several randomized controlled trials (RCTs) have been conducted to evaluate both. and demonstrated considerable reductions in ISR, target lesion and late lumen loss /vascular revascularization rates evaluated to BMS. Each was constructed from stainless steel and was thick in the struts of greater than 130m \cite{moses2003sirolimus,briguori2011novel,stettler2007outcomes}. A very late ST, while currently familiar as a potential first-generation Drug-eluting Stents (DES) complication, occurs rarely and a lot of data registries and meta-analyses consume offered comfort regarding using such methods in practice \cite{douglas2009clinical}

\subsubsection{Second-generation Drug-eluting Stents}
With the switch to metal alloys for the platform in the second-generation DES, the struts could be thinner and more flexible. The Lemus family of drugs, such as zotarolimus, enviroximes and novelist, which exhibit faster drug release and consequently earlier endothelial coverage, have been used to create new, more biocompatible polymer.

\subsection{BRS stands}
for the bioresorbable stent. Resorbing over a period of 6 months to 2 years, these stents reduce chronic inflammation over a longer period of time and promote endothelial regeneration. Reproduced with permission \cite{zong2022advances}.

\section{Complications of stenting}
Stenting complications are comparatively rare. Stents have not been associated with many complications, but there is a small chance that the body will reject the stent. Discomfort and bleeding at the puncture site where the catheter was inserted are the most common side effects.

Some people have metal allergies or sensitivities, and stents contain metal components. Stent manufacturers do not recommend using them on people sensitive to metal. Dissection of an inner layer in the coronary artery is a tiny tear that occasionally results from the procedure. The tear usually heals on its own and is not too large.

In certain circumstances, a stent is used to repair the tear. Immediate treatment is given if the tear is severe and causes blockage of arterial blood flow or bleeding out around the heart.

\section{Limitations of stenting}

In over 90 percent of patients, stenting improves blood flow and relieves symptoms, but there is a chance that symptoms will return within six months. Symptomatic restenosis can occur in the following cases: 1. About 30 percent of patients who have surgery to open a blocked artery do not receive stents. 2. Approximately 15 percent of patients with bare metal stents. 3. Less than 10 percent of patients using drug eluting stents, also known as drug eluting stents. In addition, some medical conditions, such as diabetes and continued smoking, can increase the risk of narrowing. diffuse narrowed arteries, Low-density lipoprotein (LDL) cholesterol and high blood pressure. A large blood vessel located at or near the beginning of a side branch narrows a blood vessel with numerous implanted stents.

\section{Regulatory Approval Pathways}
The 1994 FDA approval of Palmaz-Schatz stents for coronary use marked an important milestone \cite{fda1994approves}. This first bare metal stent underwent prospective clinical trials to meet regulatory requirements and demonstrate safety and efficacy. Subsequent stent approvals built on this pathway, with new generations of stents requiring comparably designed trials. First-generation drug-eluting stents gained FDA approval in 2003 to 2004 based on trials like SIRIUS \cite{moses2003sirolimus}. Approval of these novel, higher-risk devices spurred more rigorous post-marketing surveillance mandates. Costly and lengthy regulatory processes posed challenges, limiting the pace of incremental innovation. Efforts to balance safety with faster access continue to evolve.

\section{Care after the procedure}
In over 90 percent of patients, stenting improves blood flow and relieves symptoms, but there is a chance that symptoms will return within six months.

Symptomatic restenosis can occur in the following cases:

1.	About 30 percent of patients who undergo surgery to open a blocked artery do not receive stents.

2.	Approximately 15 percent of patients with bare metal stents.

3.	Less than 10 percent of patients who use drug-eluting stents, also known as drug-coated stents.

Compared to other coronary artery sites, some are more likely to re-narrow. Additionally, some medical conditions, for instance diabetes and continued smoking, might enhance the possibility of narrowing. diffusely narrowed arteries, high blood pressure, terrible cholesterol (LDL) levels that are too high.

A major blood vessel that is at the start of a side branch or close to it narrows, a blood vessel with numerous stents implanted.

\section{Preventing blood clots}
The formation of a blood clot (thrombosis) inside the stent, also known as stent thrombosis, is one of the most serious complications that can occur after the insertion of a stent. Fortunately, stent thrombosis due to administration of aspirin and other anticoagulant drugs both before and after stenting is rare. It is believed that clotting occurs when the metal of the stent comes into contact with blood components. Stent thrombosis, which cuts off the blood supply to the heart, can lead to a heart attack or even death. Even though greatest occurrences happen in the interior thirty days of stent placement, Stent thrombosis may happen as early as twenty-four hours, thirty days, or even up to a year later.

\section{When to seek help}
If any of the following events occur after stenting, seek medical help immediately:

•	Fever greater than 38°C (100.4 F).

•	You faint or feel dizzy.

•	Your pulse is not normal Start of chest pain.

•	The puncture site becomes extremely painful, swollen, warm, bleeds more than a few drops, or discharges pus.

\section{Conclusion}
The record and development stenting of coronary arteries is amongst the greatest astonishing characteristics of innovative practice of medicine. Patients requiring coronary angioplasty typically receive treatment with coronary artery stenting. Stents have eliminated the mechanical component of restenosis and acute recoil, eliminating the need for urgent bypass surgery. Even with BMS, ISR has declined due to significant advances in stent platform design. As a result of numerous efficacy and safety studies, Stents for coronary arteries are currently the preferred medication for CAD. Nevertheless, there is still thrombosis and stent restenosis a major a significant competition for modern coronary artery stents. Future interventional cardiologists may be able to use individualized, evidence-based medicine, in which the selection of a Based on stent on a patient's inherent factors, Lesion characteristics and risk profile (for thrombosis, restenosis, and bleeding).

%Bibliography
\bibliographystyle{unsrt}  
\bibliography{references}

\end{document}